%% file: prl.tex
\documentclass[notitlepage,aps,prl,reprint,superscriptaddress,nofootinbib]{revtex4-1}

\usepackage{epsfig,verbatim}
\usepackage{subfigure}
\usepackage{amsmath, amssymb, graphics}
\usepackage{color}
\usepackage{slashed}            
\usepackage{bbm}                
\usepackage{units}              
\usepackage{hyperref}           
\usepackage{xspace}             
\usepackage{enumerate}          
\usepackage{soul}
\usepackage{cancel}

\usepackage{amstext}    
\usepackage{array} 

\newcolumntype{L}[1]{>{\raggedright\let\newline\\\arraybackslash\hspace{0pt}}m{#1}}
\newcolumntype{C}[1]{>{\centering\let\newline\\\arraybackslash\hspace{0pt}}m{#1}}
\newcolumntype{R}[1]{>{\raggedleft\let\newline\\\arraybackslash\hspace{0pt}}m{#1}}

\newcommand{\mathsym}[1]{{}}

\newcommand{\eref}[1]{\eqref{#1}}

\newcommand{\ba}{\begin{eqnarray}}
\newcommand{\ea}{\end{eqnarray}}

\input{definitions.tex}

\begin{document}

\title{A new class of de Sitter vacua in String Theory Compactifications}

\author{Ana Ach\'ucarro}
\email[]{achucar@lorentz.leidenuniv.nl}
\affiliation{Instituut-Lorentz for Theoretical Physics, Universiteit Leiden, 2333 CA Leiden, The Netherlands}
\affiliation{Department of Theoretical Physics and History of Science, University of the Basque Country UPV/EHU, 48080 Bilbao, Spain}
\author{Pablo Ortiz}
\email[]{pablo.ortiz@rug.nl}
\affiliation{Van Swinderen Institute for Particle Physics and Gravity, University of Groningen, Nijenborgh 4, 9747 AG Groningen, The Netherlands}
\author{Kepa Sousa}
\email[]{kepa.sousa@ehu.es}
\affiliation{Department of Theoretical Physics and History of Science, University of the Basque Country UPV/EHU, 48080 Bilbao, Spain}

\date{\today}

\begin{abstract}
We revisit the stability of the complex structure moduli 
in the large volume regime of type-IIB flux compactifications.
We argue that when the volume is not exponentially large, such as in
K\"ahler uplifted dS vacua, the quantum  corrections to the tree-level mass spectrum can  induce tachyonic instabilities  in this sector. We discuss a  Random Matrix Theory model for the classical spectrum of the complex structure fields, and derive  a new stability bound involving the compactification volume and the (very large) number of moduli.    We also present a new class of vacua for this sector where the mass spectrum presents a finite gap,
without invoking large supersymmetric masses.
At these vacua the complex structure sector is protected from tachyonic
instabilities even at non-exponential volumes. A distinguishing feature is that
all fermions in this  sector are lighter than the
gravitino. 
\end{abstract}

%

\maketitle


 It has been known for decades that String Theory  has low energy solutions describing a four-dimensional universe with negative or zero cosmological constant, with the extra six dimensions ``compactified'' (for a review see  \cite{Baumann:2009ni,Douglas:2006es,Quevedo:2014xia,PROP:PROP200410202}).  From the four-dimensional point of view the compactified space is described by a set of fields called the moduli which describe, roughly, the size and shape of the extra dimensions.  A much harder question is whether string theory can describe a four-dimensional universe with broken supersymmetry and a positive cosmological constant, a so-called \emph{de Sitter vacuum} (dS), with a meta-stable compactification. In type IIB String Theory this question has been answered positively in a few scenarios,  the best studied being the KKLT  \cite{Kachru:2003aw} constructions, Large Vo\-lume Scenarios (LVS) \cite{Balasubramanian:2005zx,Conlon:2005} and the so-called  \emph{K\"ahler uplifted vacua} \cite{Balasubramanian:2004uy,Westphal:2006tn,Rummel:2011cd,deAlwis:2011dp,Louis:2012nb,MartinezPedrera:2012rs,Westphal:2005yz}.
The effective low energy theories des\-cri\-bing  these models
typically involve hundreds of mo\-du\-li fields, which can be divided into two classes: \emph{K\"ahler moduli} and \emph{complex structure moduli}. In addition we also have the \emph{dilaton}, whose expectation value determines   the string coupling constant.
The interactions among all these fields are given
by a complicated scalar potential, what makes a detailed perturbative
stability analysis of these vacua unfeasible except in very simplified
sce\-na\-rios.   In type-IIB flux compactifications, at the classical level,   the  scalar potential  is  induced by the presence of background fluxes (higher dimensional generalisations of electro-magnetic fields)   on the compact space \cite{Giddings:2001yu}. Due to a Dirac  condition these fluxes need to be quantised, and are therefore characterised by a set of integers.  This leading contribution of the scalar potential depends only on the dilaton and   the complex structure moduli  (for short, the \emph{complex
  structure sector}), and therefore it is necessary to take into account quantum effects   in order to fix the remaining K\"ahler moduli. 
  
 To make
the problem more tractable, it is often assumed that  the background
fluxes  provide an effective stabilisation mechanism for the complex
  structure sector, and it is not considered any further. The
consistency of this approach has been checked  for
 KKLT vacua \cite{Gallego:2008qi,Gallego:2009px,Brizi:2009nn,Choi:2004sx,deAlwis:2005tf,deAlwis:2005tg} and Large Volume Scenarios \cite{Gallego:2011jm,Cicoli:2013swa}.  Here we discuss this
matter for K\"ahler uplifted dS vacua. 

In the large volume regime of type IIB flux compactifications,  both for  LVS and K\"ahler uplifted dS vacua, the stabilisation of the K\"ahler  moduli is  a result of the competition between the leading  non-perturbative and $\alpha'$ (radiative) quantum corrections\footnote{See \cite{Berg:2007wt,Cicoli:2007xp,Cicoli:2008va} for discussions on the effect of string loop ($g_s$) corrections in these models.}. For these corrections to be under control it is necessary  that the vo\-lume of the compactification, which belongs to the K\"ahler sector, has a large expectation value compared to the string length. A large compactification volume is also essential for the consistency of the $4-$dimensional supergravity description of these models, and in particular  for the Kaluza-Klein  scale to be large compared to the supersymmetry breaking scale  \cite{Conlon:2005,Cicoli:2013swa}.
In LVS the   vacua obtained in this way   have  a negative cosmological constant (AdS), and thus  additional interactions are needed to make the  cosmological constant positive.  K\"ahler uplifted vacua are particularly interesting  because the dS vacuum is achieved without the need for  extra ingredients (matter or branes), just with an   appropriate tuning of the parameters. The downside of the latter models is that  the volume is fixed only at moderately large values, narrowing the regime of validity  of the  effective field theory.

{\bf Stability of the complex structure sector}\textemdash \, An underlying assumption of many constructions based on the scenarios above is that,
 with the right choice of  fluxes, the complex structure sector can be
stabilised at a supersymmetric configuration where the masses
of fermions and scalars are much larger than the relevant  cos\-mo\-lo\-gi\-cal energy scales. In that case, this sector can be safely integrated out, and then the attention is  focused on the stabilisation of
the lighter K\"ahler moduli, which is much trickier.  While this is a
reasonable starting point, we will argue that, at least in K\"ahler
uplifted scenarios, this assumption becomes untenable as the number of
complex structure moduli increases. We will show that this observation leads to further constraints  on the parameter space  of the model which are more restrictive than those derived from the consistency of the effective field theory.  Moreover, for very large number of
moduli --typical numbers are in excess of $\cO (100)$ -- a new class
of stable vacua emerges, in which the fermions of the complex
structure sector are all \emph {lighter} than the gravitino.

In LVS and K\"ahler uplifted vacua, the potential that stabilises the
moduli is a small deformation of the tree-level potential, with quantum
corrections suppressed by the volume $\cV $ of the compact
Calabi-Yau space \cite{Conlon:2005,Rummel:2011cd}
\be V = V_\text{tree-level} +m_{3/2}^2 \times \cO(\hat \xi/\cV),
\label{Vcorrections_intro}
\ee 
Here the parameter $\hat \xi/\cV$ characterises the magnitude of the leading quantum corrections, 
 and we have written explicitly its dependence on the volume for cla\-ri\-ty (see \cite{Rummel:2011cd,Cicoli:2007xp,Cicoli:2008va}).  The tree-level potential is positive semidefinite and is of the ``no-scale'' type: it is flat for the K\"ahler moduli leaving undetermined the expectation va\-lues of these fields, and in particular the overall volume $\cV$ and gravitino mass $m_{3/2}$. The dilaton and
complex structure moduli are stabilised at a supersymmetric
configuration that is determined by the fluxes and by the geometric and
topological properties of the compactified space \cite{Giddings:2001yu}. This configuration defines a Minkowski vacuum where, in general,  supersymmetry 
is broken by the K\"ahler moduli.

An important point is that, if we ignore quantum corrections, there is a relation between the masses of the fermions 
$m_\lambda$ (with  $\lambda$ running through the complex
structure moduli and dilaton) and the squared masses of the scalars in
the complex structure sector \cite{Achucarro:2007qa,Achucarro:2008fk,Sousa:2014qza,Marsh:2014nla} 
 $\mu_{\pm \lambda}^2$: \be \mu_{\pm
  \lambda}^2|_\text{tree-level} =( m_{3/2} \pm m_\lambda)^2.
\label{TLmasses}
\ee 
At tree level, there are no instabilities in the supersymmetric
sector, since the potential is non-negative and the no-scale vacuum is Minkowski. 
But note that, for every fermion in the supersymmetric sector
with the same mass as the gravitino, there is a massless scalar in the tree-level spectrum. The sign of the $m_{3/2}^2 \times\cO(\hat \xi / \cV)$
quantum 
 corrections is unknown so these massless scalars are not protected and can become perturbatively unstable (tachyonic, $\mu^2<0$). 
The same is true for sufficiently light scalars, to which we turn next, but  first we need to characterise the spectrum of fermion masses. 

{\bf The model}\textemdash  \,ÊThe tree-level fermion mass spectrum of the complex structure sector is determined by the geometry of the internal space and by the configuration of background fluxes. However,  the high   complexity of these theories
make a detailed calculation impractical in generic compactifications, so instead, we will follow an statistical approach.
  Intuitively, it is clear that, as the number of
complex structure moduli increases, so does the probability that there
are fermions with tree-level masses close to the gravitino mass, and
with it the
expected percentage of very light scalars that are susceptible of
becoming tachyonic by the effect of $\hat\xi / \cV$ corrections. This
intuition can be made quantitative in the framework of \emph{random matrix
theory} (RMT) \cite{mehta1991random,Aazami:2005jf,Denef:2004cf,Marsh:2011aa,Bachlechner:2012at,Long:2014fba,Brodie:2015kza}, and was confirmed in great detail in \cite{Sousa:2014qza}.

\begin{figure*}[t]
\hspace{-1cm}
{ \center \hspace{-.4cm}\includegraphics[width=0.40\textwidth]{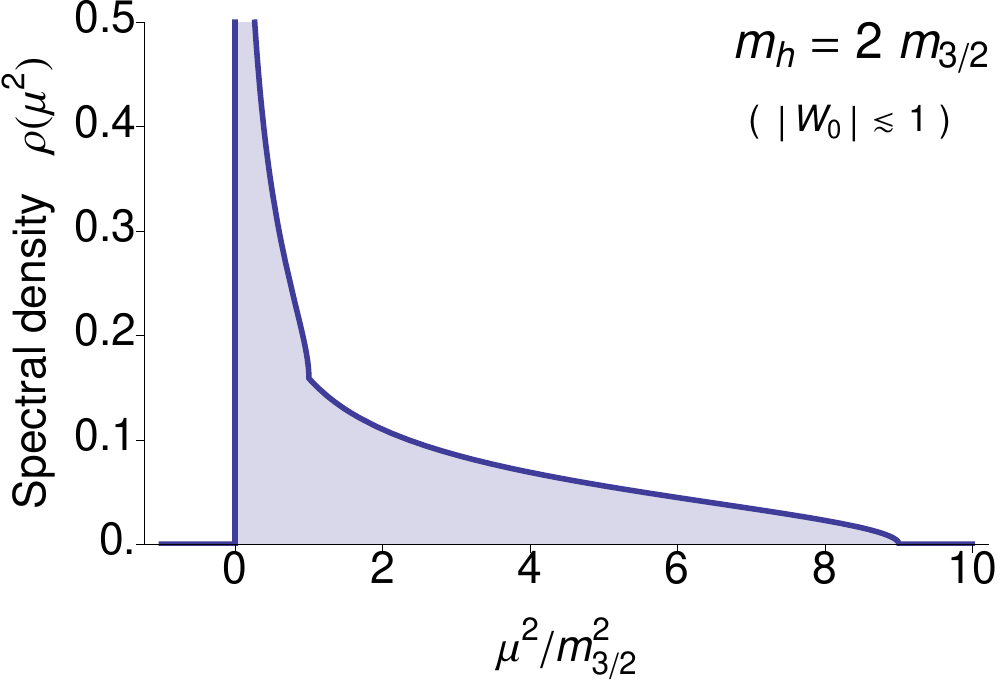}
  \hspace{1.5cm}
 \includegraphics[width=0.40\textwidth]{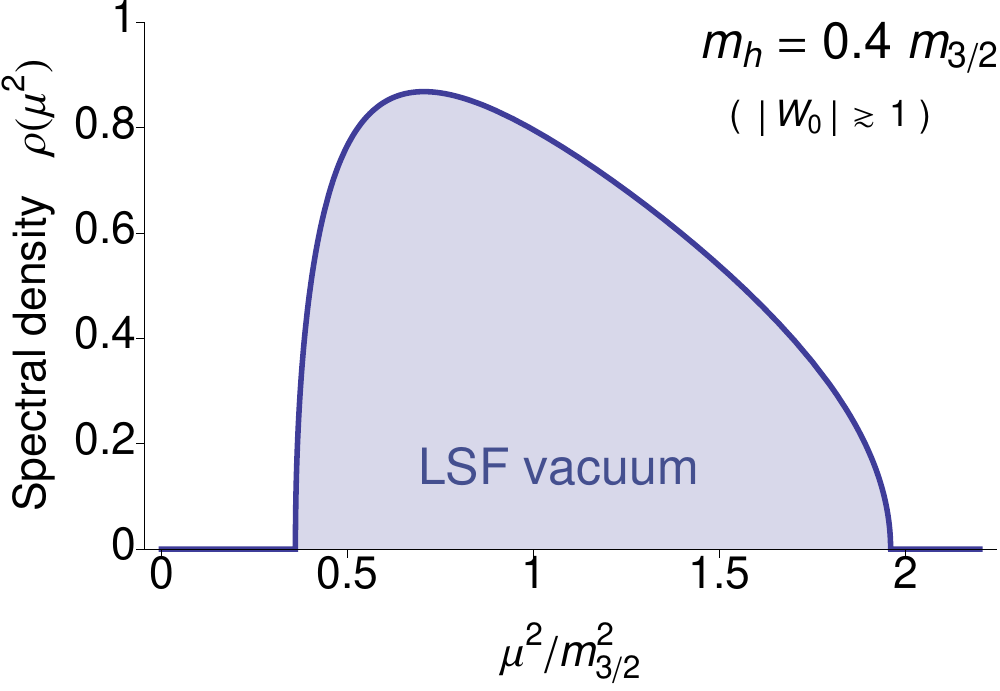}}
 \caption{Scalar mass spectrum, \eref{totaldensityM4}, of the complex
   structure sector at tree-level with a large number of fields, $N\to \infty$.  The spectrum is always
   tachyon-free, but  when the heaviest fermion is heavier than the gravitino,
   $m_h>m_{3/2}$ (left), the spectral density diverges as
   $\rho(\mu^2) \sim 1/\mu$ near $\mu=0$. By contrast,  if the heaviest fermion is lighter than the gravitino, $m_h<m_{3/2}$ (right),  the stability of the configuration is protected by a gap in the mass spectrum of size $\mu_\text{min}^2 = (m_{3/2}- m_h)^2$. }
  \label{fig1}
\end{figure*}

The idea is to  promote to random variables the entries of the  fermion mass matrix and then to characterise the spectrum of these matrices using standard techniques from RMT  \cite{Denef:2004cf,Marsh:2011aa,Bachlechner:2012at}.  The universality theorems in RMT  ensure that the result depends only mildly  on the (unknown)  distribution of the couplings for sufficiently large matrices \cite{2005math.ph...5003S,2007arXiv0707.2333H}, and therefore is insensitive  to the details of the  compactification\footnote{See \cite{Brodie:2015kza} for a recent  discussion on the applicability of random matrix theory to study  flux compactifications.}. Assuming that all complex structure moduli can be treated on equal footing, i.e. statistical isotropy in field space, the appropriate ensemble to represent the fermion mass matrix is   the Altland-Zirnbauer C$I$ matrix ensemble \cite{Denef:2004cf,Marsh:2011aa,Bachlechner:2012at}.
  Proceeding in this way, and using the relation \eqref{TLmasses}, the  authors of \cite{Sousa:2014qza} constructed   a random matrix model to characterise the tree-level scalar mass spectrum of the complex structure sector in type IIB flux compactifications. In the limit where the number of (complex) fields is large, $N\to \infty$, the spectral density $\rho(\mu^2)$ for the  tree-level  scalar masses   converges with order one probability to a particularly simple form
  \bea
\rho(\mu^2) &&  
=\frac{2 \, N m_{3/2}^2}{\pi  m_h^2 \, \mu} \Big[\nonumber\\
&&  \Theta \left(m_h^2-(m_{3/2}+ \mu)^2\right) \sqrt{m_h^2 - (m_{3/2}+ \mu)^2}+\nonumber\\
&&\Theta \left(m_h^2-(m_{3/2}- \mu)^2\right) \sqrt{m_h^2 -  (m_{3/2}- \mu)^2} \;\Big]
, \label{totaldensityM4}
\eea
where $\Theta(x)$ is the Heaviside theta function. It is important to stress that the spectral density \eqref{totaldensityM4}  is just the \emph{most likely scalar mass spectrum} predicted by the random matrix theory model. Thus, it is possible to find vacua with a different   mass spectrum, but they  occur with an exponentially suppressed probability \cite{TracyWidom,Johansson,Edelman}. 

{\bf A new class of vacua}\textemdash The spectral density \eqref{totaldensityM4} depends on two free parameters $m_h$ and $m_{3/2}$,  which re\-pre\-sent  the mass scale of the  fermions in the complex structure sector and the gravitino mass, respectively.  To be precise, the parameter $m_h$  is defined as the expectation value of the largest fermion mass $ m_h \equiv \mathbb{E}[m_{max}]$, and is related to  the flux energy scale, $m_{h} \sim M_p/\cV$, where $M_p$ stands for the Planck mass. The gravitino mass is determined by the volume  and the expectation value of the flux superpotential $W_0$,
$m_{3/2} =M_p |W_0|/\cV$ \cite{Conlon:2005}.   Figure \ref{fig1} shows the typical tree-level spectrum \eqref{totaldensityM4} of the complex structure sector. Notice the accumulation of very
light scalars in the case when the heaviest fermion is heavier than
the gravitino, ($m_h > m_{3/2}$).  By contrast, if the heaviest fermion in the complex
structure sector is lighter than the gravitino, ($m_h < m_{3/2}$), the scalar density
develops a mass gap. In the latter regime it is also  possible to find \emph{atypical vacua}, i.e. deviations from \eqref{totaldensityM4}, where the smallest scalar mass is  comparable in size to the quantum corrections, 
however the fraction of such vacua is  exponentially suppressed \cite{Sousa:2014qza}
\be
\mathbb{P}(\mu_\text{min}^2 < \frac{\hat \xi}{\cV}) \sim \rme^{-\ft{4}{3} N x^{\frac{3}{2}}},   \qquad  x =\left(1-\sqrt{\ft{\hat \xi}{\cV}}\right)^2\frac{m_{3/2}^2}{m_{h}^2} -1.
\label{PlightFields}
\ee
 The conclusion is that, for very large numbers of
moduli, $N\sim \cO(100)$, requiring the de Sitter vacua to be free from tachyonic instabilities in the supersymmetric sector  favours 
 vacua with all fermions lighter than the gravitino. We
will denote these stable de Sitter configurations ``LSF vacua'', which
stands for ``Light(er) Supersymmetric Fermions''. Note that lighter
than the gravitino does not necessarily mean light; the actual fermion
masses can easily be in the Grand Unification scale as long as the
gravitino is even heavier\footnote{If $m_h\sim \cO(m_{3/2})$, 
  the smallest fermion mass is of order $m_{3/2}/N$ \cite{Sousa:2014qza,Bachlechner:2012at}.  In K\"ahler uplifted vacua  $m_{3/2}$ is typically  of the order of the GUT scale \cite{Westphal:2006tn}, leading to  $m_{min}\sim M_{GUT} \cdot10^{-2}$.}.

\begin{figure}[t]
 \centering \includegraphics[width=0.43\textwidth]{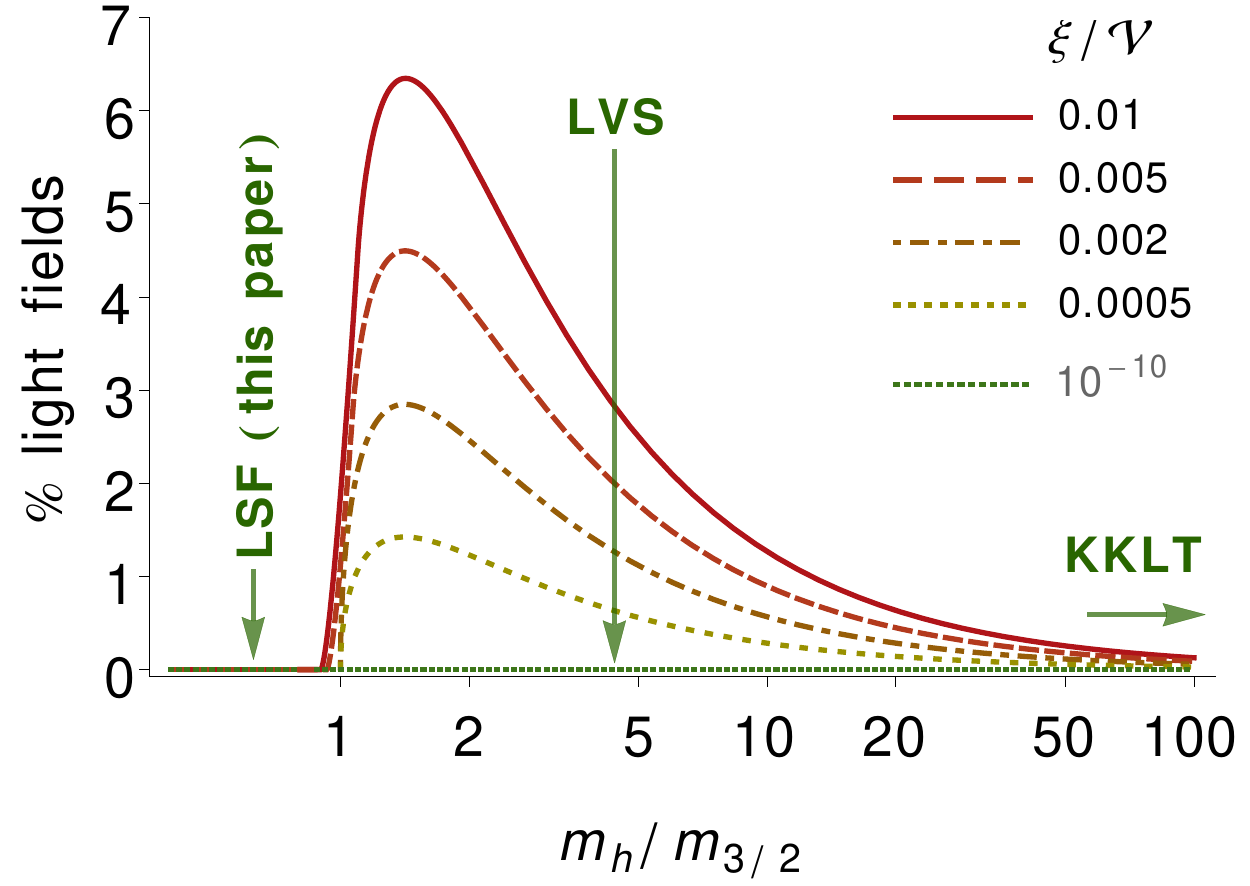}
    \caption{Percentage of (real) scalars  in the complex structure sector with  tree-level masses  smaller than the size of the leading quantum corrections, $\mu^2 \le m_{3/2}^2 \cdot \hat\xi/{\cal V}$. The horizontal axis  represents  the typical  mass scale  in  this sector, $m_h$.
The spectrum of perturbations of the LSF vacua ($m_h< m_{3/2}$),  contains no light scalar modes at tree-level. Stability is also ensured if there is a large hierarchy  between the masses of the supersymmetric complex structure sector and the supersymmetry breaking scale, $m_h\gg m_{3/2}$ (KKLT regime), or an exponentially large volume, $\hat\xi/{\cal V} \sim 10^{-10}$  (LVS).  
}
 \label{fig2}
\end{figure}

{\bf Comparison with KKLT and LVS regimes}\textemdash The KKLT scenario corresponds to fine-tuning the fluxes so that the complex structure moduli have large masses compared with the supersymmetry breaking scale, which is set by the gravitino mass, that is $m_h \gg m_{3/2}$  ($|W_0| << 1$) \cite{Gallego:2008qi,Gallego:2009px}. In this regime  the stability of this sector is guaranteed since the tree-level masses are large compared to the contributions induced by quantum effects. In LVS and in K\"ahler uplifted vacua the absence of fine-tuning of $W_0$
 implies that the fermions typically have masses comparable (but not necessarily close) to the
gravitino mass, so that generically we have
$ \mu_{\pm \lambda}^2\sim \cO(m_{3/2}^2)$
\cite{Conlon:2005ki,Gallego:2011jm,Baumann:2009ni}.
The corrections to the tree-level spectrum can still be consistently
neglected in this setting as long as the volume of the compactification is
exponentially large, and thus the corrections are tiny,
$\hat \xi/\cV \sim 10^{-10}$.
  For K\"ahler uplifted
vacua this is no longer true. Since the volume is not exponentially
large, typically we have $\hat \xi/\cV \sim 10^{-2}-10^{-4}$ \cite{Westphal:2006tn,Rummel:2011cd,Louis:2012nb},
implying that the corrections in \eqref{Vcorrections_intro} could in principle
induce tachyonic instabilities if some of the complex structure moduli
are sufficiently light at tree-level, \be \mu_{\pm
  \lambda}^2|_\text{tree-level}\lesssim m_{3/2}^2 \, \times \,
\cO\left(10^{-2}-10^{-4}\right).
\label{danger}
\ee
 Figure \ref{fig2} shows the percentage of scalar moduli  estimated
using \eqref{totaldensityM4} that are light enough
to be destabilised by the quantum corrections, for a range of values of $\hat \xi /\cV$. 
Note that, for moderately large volumes $\hat \xi /\cV\sim 0.01$, this fraction can rise up to a $6-7\%$, with the maximum occurring at $m_h\approx \, \sqrt{2} \, m_{3/2}$. 

Requiring that the number of light fields,  $N_\text{light}$, is less than one irrespective  of the details of the stabilisation of the complex structure sector, i.e. regardless of the value of the mass parameter $m_h$, we find a bound for the size of the $\alpha'$ corrections  
\be
\max\{N_\text{light}\}\approx \frac{4 N}{\pi}\sqrt{\frac{\hat \xi}{\cV}}\ll1 \quad\Longrightarrow \quad  \frac{\hat \xi}{\cV} \ll  \frac{\pi^2}{16 N^2}.
\label{bound}\ee
Note that in a generic compactification with hundreds of complex structure moduli this bound  is much more restrictive than just requiring the $\alpha'$ corrections to be small, $\hat \xi/\cV\ll1$. This remark is particularly relevant  for dS solutions and inflationary models built with the method of K\"ahler uplifting that do not satisfy the constraint    \eqref{bound} (see examples in \cite{Westphal:2006tn,Rummel:2011cd,MartinezPedrera:2012rs,Westphal:2005yz,Ben-Dayan:2013fva,Pedro:2013qga}), as this signals the possible  presence of tachyonic instabilities.  Other models which could be affected by the same issue are those based on LVS vacua where the volume  is only moderately large   \cite{Rummel:2013yta,Sumitomo:2013vla,Maharana:2015saa,Cicoli:2015wja}.  
In all these constructions one could  still search for atypical vacua where all fields in the complex structure sector are much heavier than the gravitino, as in KKLT scenarios.  However the probability of such vacua  is  exponentially suppressed as, without  fine-tuning $W_0$,  the parameters satisfy $m_h \sim m_{3/2}$, and thus  \cite{Bachlechner:2012at,Sousa:2014qza} 
\vspace{-.3cm}
\be
\mathbb{P}(\mu^2_{\text{min}} \ge m_{3/2}^2) \sim \rme^{- \frac{8 m_{3/2}^2 }{m_h^2} N^2} \ll1.
\ee
By contrast, LSF vacua,  where all fermions  are lighter than the gravitino,  occur with  probability of order one when  $m_h\lesssim m_{3/2}$ ($|W_0| \gtrsim 1$), and thus  they are a  more natural configuration to stabilise the complex structure sector  in this regime. In figure \ref{fig2} it can be seen that when LSF vacua become dominant, the typical spectrum contains no light fields, a direct consequence of the appea\-rance of the  mass gap. Other scenarios which satisfy   constraint \eqref{bound} are \cite{Braun:2015pza,Westphal:2005yz,deAlwis:2011dp,Sumitomo:2013vla}.

 {\bf Discussion}\textemdash \,  Having established that LSF vacua are stable, the next question is how
to find them.  Reference
\cite{attractors} provides a systematic way of looking for LSF vacua by
looking in the vicinity of configurations in which all fermions in the
supersymmetric sector are \emph{massless}. The Massless Fermion Limit (MFL)
is not always realised at a physical vacuum, because the massless
condition may require non-integer values of the fluxes that are not
actually realised. However, it provides the ``lamppost'' near which
actual stable vacua may be found.

This brings us to another important point. 
The explicit examples of K\"ahler uplifted vacua constructed to date
\cite{Westphal:2006tn,Louis:2012nb,MartinezPedrera:2012rs} have been found in models
consistent with the supersymmetric truncation of a large sector of the
moduli fields \cite{Warner:1983du,Denef:2004dm,DeWolfe:2004ns,Cicoli:2013cha,Giryavets:2003vd,Louis:2012nb,He:2013yk,BlancoPillado:2012cb}.  This can be achieved by considering special
points of the moduli space, for ins\-tan\-ce fixed points of global
symmetries of the moduli space metric, where the majority of the
fields can be fixed at a supersymmetric configuration. By this
procedure it is possible to obtain a reduced theory involving, in
addition to the K\"ahler moduli, a small fraction of the complex
structure fields and the dilaton so that a detailed stability analysis
is possible. In particular, in the examples discussed in
\cite{Westphal:2006tn,Louis:2012nb} the complex structure moduli
surviving the truncation were fixed at vacua with large supersymmetric
masses, i.e. $m_\lambda \gg m_{3/2}$, that is, imposing a large
hierarchy between the masses of these fields and the supersymmetry
breaking scale. This method ensures the stability of the field
configuration in the reduced theory, however it cannot guarantee that
the truncated fields are fixed at minima of the potential and, for
this reason, neither does it guarantee the consistency of this
reduction.  It is therefore crucial to understand under what
conditions it is possible to ensure the stability of the full set of
moduli fields, \emph{including the truncated ones}.

In paper \cite{attractors} it is also shown that, when the fraction of
complex structure fields surviving the truncation are stabilised at
the MFL of a critical point, then \emph{all} the complex structure
fields (including the truncated ones) and the dilaton have a mass
equal to $m_{3/2}$ at tree-level, i.e. the full sector is also at the
MFL of the vacuum.

Given that it is not feasible to check the stability of hundreds of
supersymmetric moduli --except perhaps in very special cases--, we
would like to suggest a compromise: apply the usual analytic and
numerical techniques to check stability of the surviving low-energy
sector (ty\-pi\-cal\-ly, the K\"ahler moduli and the complex moduli that sit
at points of enhanced symmetry) and supplement these with the use of random matrix theory techniques
to assess the stability of the truncated moduli that do not appear in
the low energy description. Here we made use of the random matrix
theory model presented in \cite{Sousa:2014qza} to
characterise the mass spectrum of the complex structure sector. Our
conclusion --in line with our previous work in \cite{Achucarro:2007qa,Achucarro:2008fk,Sousa:2014qza}-- is that,
in compactifications where the number of complex structure moduli is
very large, there is a class of stable flux configurations, not previously
con\-si\-der\-ed, in which all fermions of the supersymmetric sector
--including truncated ones-- are lighter than the gravitino.
\newpage
\emph{Acknowledgements}\textemdash \,  We thank J.J. Blanco-Pillado, J. Urrestilla, J. Frazer,  F. Denef,  M. Rummel, B. Vercnocke, T. Wrase,  and specially David Marsh for many  helpful discussions.
The authors acknowledge financial support from the  Dutch Organization for Fundamental Research in Matter (F.O.M.)  under the program ``Observing the Big Bang'', the Netherlands Organisation for Scientific Research (N.W.O.) under the ``Gravitation'' program, the  Spanish Consolider-Ingenio 2010 program CPAN CDS2007-00042, the Consolider EPI CSD2010-00064, 
   the Basque Government (IT-559-10) and the Spanish Ministry of Science grant (FPA 2012-34456). P.O.   acknowledges the hospitality of the University of the Basque Country UPV/EHU and the University of Leiden during the completion of this work. K.S. would like to thank to  the Department of Physics of the University of Oxford for its hospitality.

 \bibliography{prl.bib}

\end{document}

%% file: definitions.tex
\newcommand{\be}{\begin{equation}}
\newcommand{\ee}{\end{equation}}
\newcommand{\bea}{\begin{eqnarray}} 
\newcommand{\eea}{\end{eqnarray}}
\newcommand{\ft}[2]{{\textstyle\frac{#1}{#2}}}

\def\rme{{\mathrm e}}

\newsavebox{\uuunit}
\sbox{\uuunit}
    {\setlength{\unitlength}{0.825em}
     \begin{picture}(0.6,0.7)
        \thinlines
        \put(0,0){\line(1,0){0.5}}
        \put(0.15,0){\line(0,1){0.7}}
        \put(0.35,0){\line(0,1){0.8}}
       \multiput(0.3,0.8)(-0.04,-0.02){12}{\rule{0.5pt}{0.5pt}}
     \end {picture}}

   \def\cO{{\cal O}}

    \def\cV{{\cal V}}

\def\atH0{|_{H_{0}}}
\def\AtH0{\bigg|_{H_{0}}}

\csname @addtoreset\endcsname{equation}{section}
